# The Certification of ATLAS Thin Gap Chambers Produced in Israel and China

E. Etzion, Y. Benhammou, J. Ginzburg, M. Ishino, L. Levinson, G. Mikenberg,
N. Panikashvili, D. Primor, Y. Rozen, V. Smakhtin and S. Tarem,

*Abstract*--Thin gap chambers (TGCs) are used for the muon trigger system in the forward region of the LHC experiment ATLAS. A TGC consists of a plane of closely spaced wires maintained at positive high voltage, sandwiched between resistive grounded cathode planes with an anode wire to cathode plane gap distance smaller than the wire-to-wire spacing. The TGCs are expected to provide a trigger signal within 25 ns of the bunch spacing of the LHC accelerator, with an efficiency exceeding 95%, while exposed to an effective photon and neutron background ranging from 30 to 500 Hz/cm$^2$. About 2,500 out of the 3,600 ATLAS TGCs are being produced at the Weizmann institute in Israel, and in Shandong University in China. Once installed in the ATLAS detector the TGCs will be inaccessible. A vigorous production quality control program is therefore implemented at the production sites. Furthermore, after chamber completion, a thorough program of quality assurance is implemented to ensure the efficient performance of the chambers during more than ten years of operation in the LHC high rate environment. This program consists of a detailed mapping of the detectors response using cosmic rays, as well as checking the chambers behavior using a high rate radiation source. An aging test performed on five chambers in a serial gas connection is presented. Finally the results of the chambers certification tests performed at CERN before the installation in ATLAS are described.

Manuscript received October 30, 2004. This work is supported by the Israel Science Foundation and the German Israeli Foundation.
  E. Etzion, Y. Benhammou, J. Ginzburg and D. Primor are with the School of Physics and Astronomy, Raymond and Beverly Sackler Faculty of Exact Sciences, Tel Aviv University, Tel Aviv 69978, Israel (telephone: +41764870750, e-mail: erez.etzion@cern.ch).
  N. Panikashvili Y. Rozen and S. Tarem are with Physics Department, Technion, Haifa, Israel.
  L. Levinson, G. Mikenberg and V. Smakhtin are with the Particles Physics Department, Weizmann Institute of Science, Rehovot 76100, Israel.
  M. Ishino is with ICEPP, University of Tokyo, 7-3-1 Hongo, Bunkyo-ku,Tokyo, Japan.

## I. INTRODUCTION

The two beams at the proton-proton Large Hadron Collider (LHC) will reach the energy of 7 TeV, the highest energy ever achieved in a particle accelerator. Bunches of protons will collide inside the ATLAS detector at a rate of 40 MHz namely every 25 ns, reaching a design luminosity of $10^{34}$ cm$^{-2}$s$^{-1}$. The accelerator and the ATLAS detector which are currently built at the European Center for Nuclear Research (CERN) are scheduled to start operating in the year 2007. The extremely high event rate requires a uniquely efficient and robust triggering system. The trigger system is aimed at reducing the event rate several orders of magnitude from the level of $10^9$ Hz to about 100 Hz without risking the potential detection of rare new physics processes, possibly new phenomena or new objects. The ATLAS Muon spectrometer [1] will have two types of detectors which are dedicated to trigger on muon particles based on their transverse momentum (Pt). These detectors are required to demonstrate fast detection with a very small dead time and to provide moderate Pt measurement and bunch identification. These chambers are also designed to provide an azimuthal coordinate measurement as an input to the tracking system. The muon trigger will be derived from three trigger stations formed of Resistive Plate Chambers (RPC) which are very fast detectors positioned in the ATLAS barrel region and TGCs [2] which are moderately fast detectors but can operate at higher noise rate environment as expected in the ATLAS two end-cap regions. The Muon end-cap trigger consists of seven layers of TGCs: one triplet (each consists of three wire planes) and two TGC doublets (two wire planes). The chambers are mounted on big wheels in the end-cap region about 14 m away from the interaction point. A TGC consists of a plane of closely spaced wires maintained at a positive high voltage (HV), sandwiched between resistive grounded cathode planes. A TGC is characterized by an anode wire to cathode plane gap distance of 1.4 mm which is smaller than the 1.8 mm wire-to-wire spacing. The sizes of TGC units depend on their position and vary between 1.31 $m^2$ to 2.27 $m^2$.

The TGCs are well suited for triggering the high transverse momentum muons at the LHC as they operate in a saturated proportional mode, giving fast signals with a typical rise time below 5 ns. The saturated proportional mode is gained under a strong electric field obtained with 50 microns gold plated tungsten wires and a very narrow gas gap. The shower amplifications and quenching are optimized with a mixture of 45% n-Pentane and 55% $CO_2$. This operation mode leads to strong signals with reduced Landau tails and a high signal to noise ratio. To form a trigger signal, between 4 to 20 anode wires are grouped together and fed to a common readout channel. The wire and strip signals emerging from the TGCs are fed into a two-stage amplifier in an Amplifier Shape Discriminator (ASD) circuit [3]. Four such circuits are built into a single ASD chip and four ASD chips are incorporated into an ASD Board.

Three production lines build the TGCs for ATLAS. The construction of the 3,600 chambers is shared between Weizmann Institute of Science (Israel) building 2,160 chambers, KEK (Japan) building 1,056 chambers and Shandong University (China) building 384 chambers. The chambers built in Japan are tested in a testing facility in Kobe University [4], the chambers built in Israel and China are tested in the two cosmic ray (CR) hodoscopes constructed for that purpose in the Technion and Tel Aviv University [5].

## II. TESTS IN THE PRODUCTION LINE

After the installation in the ATLAS detector, during operation the TGCs will be inaccessible. This dictated a vigorous Quality Control (QC) program which starts at the production sites. The production tooling and the procedures are all defined in details in the construction manual document [6]. The QC starts with a pre-construction stage where the uniformity of the material to be used for the chambers production is examined. This includes a verification that the gap boards' thickness is within their nominal value of 1.4 +/- 0.1 mm, and the deviation from flatness is less than 0.2 mm. The strips are visually and electrically tested looking for shorts or cuts. The frames and the wire supports are both tested under magnifying glass while painting the edges with an epoxy layer.
The QC study continues as part of the production cycle where for example it is verified that the planarity of the chamber is better than 0.2 mm after gluing the chamber frame.
Before closing a unit, every ½ detector is tested to hold voltage of 2,500 volts with current below 20 micro amperes.
In the next stage the chambers are tested to hold 2.8 kV while running with $CO_2$ and 3.2 kV running with $CO_2$ n-Pentane mixture.

After the unit is closed each module goes through a set of tests which check its stable operation below current limits running at the operational voltage, operation of all the channels in the detectors, and its gas tightness. The test procedure includes the following step:

- Stable operation is first tested with $CO_2$ at 2,900V. The test continues with the nominal $CO_2$ n-Pentane mix at 3,200 V (300 V above the nominal operation voltage) for 24 hours. The detector is required to run with a current below three micro amperes.
- No more than three disconnected channels per chamber are allowed. Nevertheless for large wire groups even this criterion can already badly impact the inefficient area; hence even detectors with two unconnected wire groups may fail the criteria of efficiency as measured later in the cosmic stands.
- Gas tightness is measured for both the chambers gas gap and the envelope, which for safety reasons is to be operated with a continuous flow of $CO_2$. During the test the chamber is operated with two times the operating overpressure. The detector pressure drop is measured and is required to be less than 0.05 cc/minute. The same technique is applied to the $CO_2$ envelope.

All the checks and intermediate measurements are recorded in the production Object Oriented database [7] in Weizmann for chamber made in Israel and China or an Access database [8] for the chambers made in Japan.

## III. RADIATION TEST AND AGING

Figure 1 presents a side view of the TGCs in one quarter of the ATLAS detector. The TGC trigger chambers are orders in two layers of doublets and one triplet in the middle station between 13 to 15 meters from the interaction point. Additional TGCs doublets which deliver additional coordinate measurement to the tracking system are located in the inner station about 7 meters from the interaction point. T1-T11 are the names of the TGCs different types of units. Figure 2 presents the expected total count rate of background in $Hz/cm^2$ and its dependence on the pseudorapidity for both the inner and the middle station. As seen in the figure the background is expected to reach between about 20 $Hz/cm^2$ in low $\eta$ of the middle station to above 100 $Hz/cm^2$ for high $\eta$ in the inner station.
To be able to achieve the required performance, without failures under radiation, all the Israeli and China made TGCs are exposed in the radiation facility of Weizmann Institute of Science biology Laboratories to $^{60}$Co-$\gamma$ high radiation dose [9]. The Radiation test is sensitive to the quality and homogeneity of the graphite and its grounding contact. The detectors are irradiated with high radiation gamma-rays (300 $Hz/cm^2$ which is about three times higher than the maximal expected radiation rate for the TGCs in the inner station) for 20-30 minutes. This is a useful test to detect glue spots left inside the gas gaps, which could lead to high local currents.
A comparison of the leak current and the noise rate is done between before, during and after the irradiation. During the irradiation, the chamber should be well above the plateau, namely a decrease in voltage by 100 V, should lead to a rate decrease of <10%, while the gain, defined by I/rate, should be larger than 5 Pc. As for the current and noise rate during the irradiation there should be an increase in the current and noise rate.

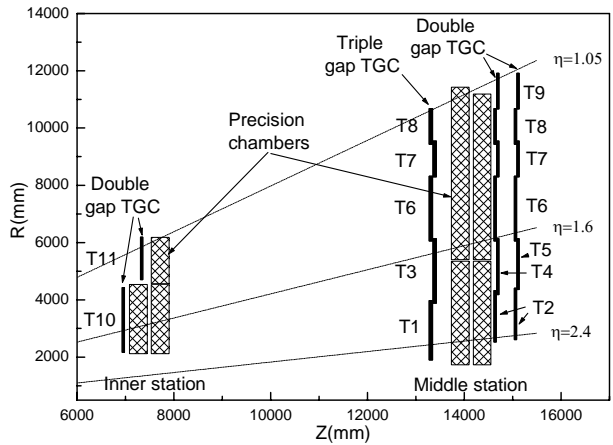

Figure 1- A longitudinal view of the TGC System in the ATLAS experimnet

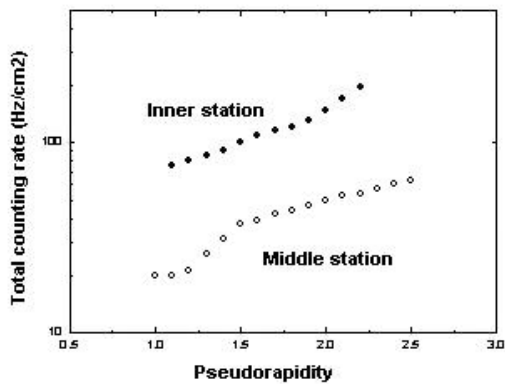

Figure 2 - A logaritmic view of the simulated total background in the inner and middle stations as a function of the pseudorapidity

After turning off the source, the chamber current should go back to its original value to within 1 micro ampere, and the differential rate increase should not be larger than 0.1 Hz/cm$^2$. This is indeed the case for good chambers, however for several problematic chambers the current and the noise rate returned to values higher than their initial values. An example of a mull-functioning detector is given in Figure 3. Here there is no current before irradiation, it jumps to the level of 12-15 micro amperes and when turning off the chamber it decreases but doesn't go to zero but stay at the level of 2 micro amperes. It only returns to zero after turning off the HV supply. For a good chamber we get dI equal to zero.

Problematic chambers went through a two stages treatment: First stage was to repeat the test with a rate ~3 times higher (1 kHz/cm ), to burn possible dust inside the gas gap.

Usually it solved the problem of noise and high currents. For some chambers constructed in the early construction stage, the first wire was 2 mm away from the frame, and this caused in some cases discharges to small pieces of glue inside the chamber volume. In those cases the first wire group was disconnected.

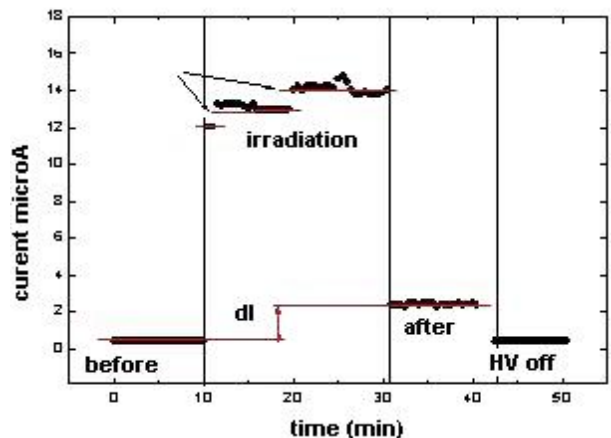

Figure 3 - Leak current distribution during irradiation. Shown here for a problematic chamber.

1543 have already been tested in this scheme (697 of the T8 type, 451-T6, 279-T3, 92-T9, 18-T1 and 6-T11). Figure 1 presents the current summary of the irradiation studies. The logarithmic view histogram shows that in a large majority of the chambers the difference between the current before and after irradiation is well below the dI<1 micro ampere requirement.

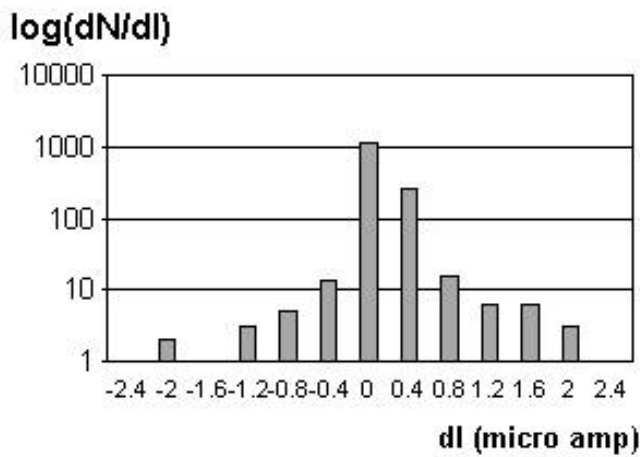

Figure 4 – A Summary histogram of the irradiation tests, a logarithmic plot of dI (micro amperes) distribution.

ATLAS is expected to run for long time in the very high radiation background environment of the experimental hall. The TGCs will be exposed to up to a few hundred Hz/cm$^2$ for about 3,000 active hours a year for at least 10 years. A long term gamma irradiation test was held in order to study aging effects on the TGC operation using the source described above. A set of chambers was irradiated during operation in order to determine whether irradiation can impact the TGC performance by destroying the large molecules of the n-Pentane gas. A stand of five T8 type TGC doublets was used for this study.

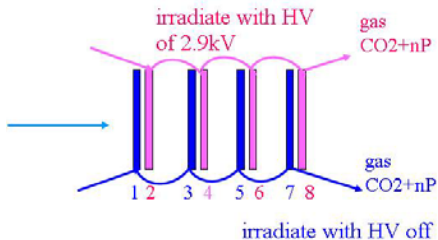

*Figure 5 - The aging test chambers setup.*

Figure 5 shows a schematic view of the detectors in the test. Five doublets (10 detectors) took part. A coincidence between hits detected in the first and the last chambers (not shown in the figure) was used as a trigger and reference for the detection in the other chambers. In the test sample four chambers (1, 3, 5, 7) were chained in one gas line supplying the nominal mixture of $CO_2$ and n-Pentane. They were used as a reference operated without HV. The other four chambers (2, 4, 6, 8) where chained to a second line of mixed gas however they were operated at 2.9 kV. The whole setup was irradiated with gamma-rays from $^{60}$Co-$\gamma$ providing a rate of, 1,200 Hz/cm$^2$ for 300 hours in 40 days. In that way the chambers accumulated charge of 1.5 mC/cm$^2$, a level which is equivalent to three years of operation in the ATLAS rate.

Some of the results of the aging tests are plotted in Figure 6. The ratio on the top is of the signals in the detectors which were irradiated and operated at HV of 2.9kV over those running without HV. The second ratio shows the rate of signals in the last two over the first two chambers in the gas chain.

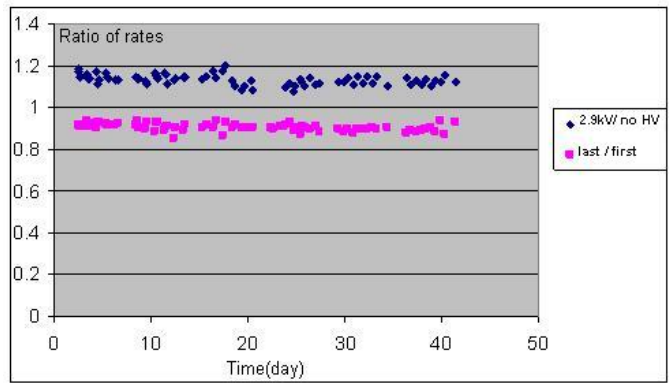

*Figure 6 - Aging results. The blue points on the top present the ratios of cosmic rates in the chambers irradiated with HV over those irradiated without HV. Below (pink points) are for the ratio of the two last chambers in the gas chain over the first two chambers.*

No degradation due to irradiation or degradation due to the position in the gas chain was found during the whole operation period of the test.

A study of the operation at two different readout thresholds along the irradiation period showed no change in the radiation spectrum due to irradiation.

IV. COSMIC RAY TESTBENCH

CR hodoscopes are used to measure the time response and the detection efficiency of the TGCs [5]. Two similar systems one at the Technion and one at Tel Aviv University are testing all the chambers which were built in Israel of China. Up to eight TGC units (doublets or triplets) which are tested in parallel are interleaved between two precision chambers (PRC) and two layers of scintillators. The indication to a muon crossing the PRC, and therefore the tested stack, is provided by coincidence between the two scintillator planes, one above the upper PRC and one below the lower PRC. For calculating the hit position we read out the signals from the PRCs. These are special rectangular shape TGCs, which consists of two perpendicular layers of strips and one layer of HV wires between them. Each strip is 3.6 mm wide. The signals from the two layers are latched and read sequentially providing the X and Y coordinates of the hit position. The hit position is derived by using the charge distribution between adjacent strips, this gives us a resolution of a few hundred microns. A straight line interpolation between the two PRCs is used to predict the hit expected position in the tested TGCs. The results of the test are detailed efficiency map with a position resolution of 1x1 cm$^2$. A good chamber is required to have at least 95% of its active area more than 95% efficient with a time response of less than 25 ns. Figure 7 is an example of an efficiency map of one of the chambers. In the figure the strips efficiency of a T8 type chamber. One can clearly see in the plot the borders of the chamber as well as its five support lines and 50 wires support buttons. The results of the CR efficiency tests are all summarized and stored in local databases at the tests sites. The laboratories at the Technion and Tel Aviv have tested already more than 280 T8 Units, 280 T6 type units and 24 T3 units (both doublets and triplets). Two shipments of T9-type chambers with 170 units which were produced in Shandong University arrived to Israel and more than half of them passed the cosmic efficiency and the irradiation tests.

V. CERTIFICATION AT CERN

Following a damage found in two chambers delivered by air from Israel to CERN, all other chambers were shipped over the sea carried on a track specially designed to soften the shocks on the way. The chambers are delivered in packages of about 30 chambers each with 10-15 packages on a track. Upon their arrival to CERN, all the chambers go through a set of final tests. These tests contain search for cut wires, gas tightness, HV stability and signal tests [10].

To assure there are no shorts due to a wire cut on the way, all the HV connectors are tested with a Mega ohm tester, requiring that the resistance of each HV line to be higher than $2\,G\Omega$.

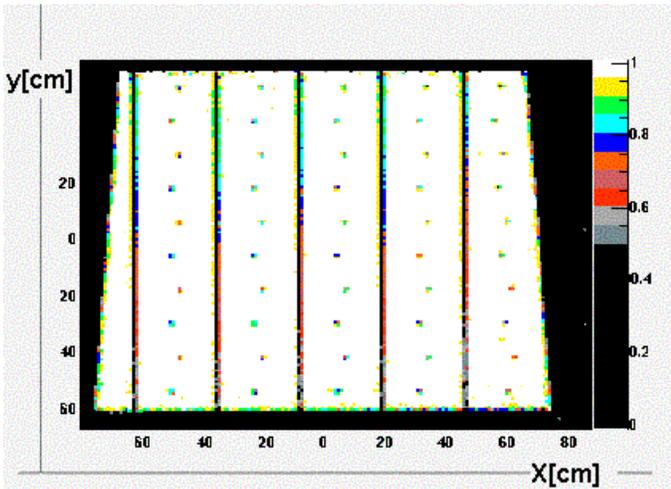

*Figure 7 – An efficiency map of one of the chamber strips. The bin efficiency level is color coded with white standing for efficiency>95% and black for efficiency below 50%. The scale is given on the right side of the plot.*

To verify that there are no gas leaks each chamber is filled with $CO_2$ at an initial pressure of 400 Pascal. It is required that the pressure drop will be less than 40 Pascal in 5 minutes.

A similar test is applied also to all the $CO_2$ channels on the envelopes of the units.

The stability under HV and the signal tests are done on a set of four packages together. The packages are moved to a close room which is equipped and approved for flammable gas use.

An external gas system supplies the test room with $CO_2$ or $CO_2$ n-Pentane mixed gas. In the room the gas is distributed to the chambers chained in serial gas lines of about 10 chambers in a row. This is similar to the gas distribution planned for the ATLAS experiment with a flow rate of 2 liter/hour in each line. Three units of CAEN SY127 supply power to 120 HV channels used in the test. The HV is remotely controlled and continuously recorded. In Figure 8 a view of the test room with 120 chambers (3% of ATLAS TGCs) in operational mode.

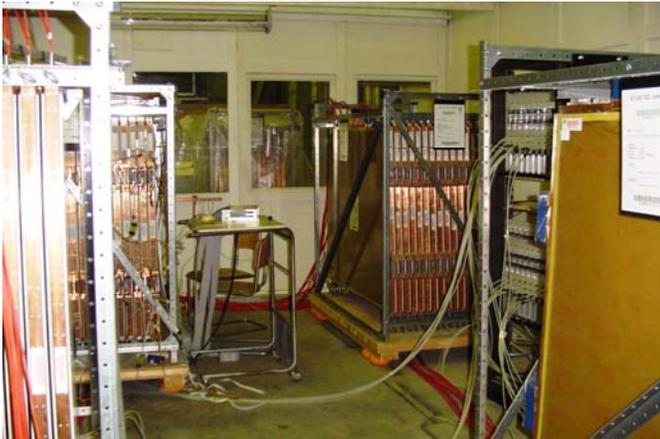

*Figure 8 - A picture of the test room with 120 chambers connected to Gas, HV and readout.*

The four packages go through a test cycle of about a month in this test room. The chambers are flushed with $CO_2$ for four days. After four days they are provided with 2.7 kV and in case no problems occur the HV is turned off and gas supply is switched to the operational $CO_2$ and n-Pentane gas mixture. It takes a week to reach a homogenous gas mixture inside the chamber which enables to ramp the HV to 2.9 kV.

The chambers are operated at that value for two weeks recording the stability of the current in the chambers. During that period the rate of signals produced by CR muon is measured.

Two systems to accumulate a hit count for each channel were developed; one at the Weizmann Institute and the second at Shandong University. We use the two to confirm that all the channels are operational and run at a low noise level. The systems supply the low voltage required by the TGC-ASD readout cards and read their output signals in a self trigger mode.

There are nominal threshold values which should suit all the strip lines and wire lines; however for the exception cases the Israeli system (TGC-Lite) can supply different threshold values to different readout cards

The summary data is written to an output file tracked by the certification database (DB).

All the data collected during the production, the radiation tests and the CR tests is imported to a unique Oracle certification DB. This DB, which has a web interface access, is used also to store the results of the QC tests at CERN and track the test status. Chambers that successfully pass this stage of QC are certified for installation in the ATLAS detector.

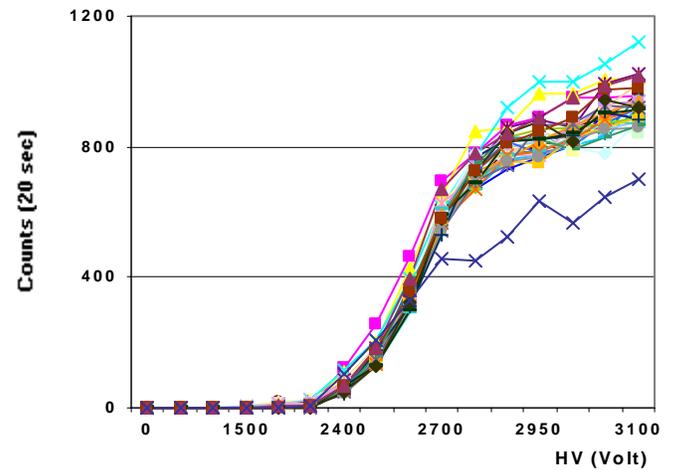

*Figure 9 - Plateu curve, Counts in 20 sec as a function of the HV supplied to the chamber. Each line in the figure represents one (of 32) strip channels of a T8 detector.*

## VI. SUMMARY

The production of the TGCs is close to its end, more than 85% of the chambers were produced, 55% of them were tested in the cosmic stands and in the irradiation tests.

The CR and irradiation tests are critical stages in the QC procedure of the chambers which successfully detects problematic chambers, providing an efficiency map for each chamber and suggests operational parameters such as voltage and readout thresholds.

Extensive aging tests have been performed on final detectors using the 2kCi $^{60}$Co-$\gamma$ source available at the Weizmann Institute. The test aimed at evaluating whether there are any degradation effects caused by the destruction of n-Pentane molecules under irradiation. The performance was measured along a set of chambers connected in a serial line. The study did not detect any degradation problems related to irradiation.

630 doublets and triplets (43% of ATLAS TGCs) have already been delivered to CERN where they are required to pass the final QC procedure. No cut wire was found in any of the chambers. A gas leak beyond the allowed level was found (and fixed) in two of the chambers, were 14 had leaks in their $CO_2$ envelope. 850 chambers which are about 24% of the ATLAS TGCs went through the whole testing procedure. Two chambers failed in the tests where 24 of them had one or two noisy channels. The results of the tests are stored in the ORACLE Certification DB which is accessible via standard WEB tools.

## VII. ACKNOWLEDGMENT

We thank all our colleagues from the TGC groups in China, Japan and Israel.